\documentclass{jfm}
\usepackage{graphicx}
\usepackage{epstopdf}
\usepackage{float}

\usepackage{amsmath}
\usepackage[colorlinks, citecolor=blue]{hyperref}
\usepackage[dvipsnames]{xcolor}
\usepackage{comment}
\usepackage[round, comma]{natbib}
\usepackage{microtype}
\usepackage{enumitem}

\usepackage{lineno}
\usepackage{placeins}
\usepackage[normalem]{ulem}

\usepackage{siunitx} 

\newcommand*\coltext{\textcolor{Black}}

\shorttitle{Capillary evaporation of binary liquids}
\shortauthor{L. Thayyil Raju et al.}

\title{Evaporation of binary liquids from a capillary tube}

\author{Lijun Thayyil Raju\aff{1}
  \corresp{\email{l.thayyilraju@utwente.nl}},
  Christian Diddens\aff{1},
  Javier Rodr\'{\i}guez-Rodr\'{i}guez\aff{2},
  Marjolein N. van der Linden{\aff{1}$^{,}$\aff{3}},
  Xuehua Zhang{\aff{1}$^{,}$\aff{4}},
 Detlef Lohse{\aff{1}$^{,}$\aff{5}}
 \corresp{\email{d.lohse@utwente.nl}},
 \and Uddalok Sen{\aff{1}$^{,}$\aff{6}}
  \corresp{\email{uddalok.sen@wur.nl}}}

\affiliation{\aff{1}Physics of Fluids Group, Max Planck Center Twente for Complex Fluid Dynamics, Department of Science and Technology, and J. M. Burgers Centre for Fluid Dynamics, University of Twente,  P. O. Box 217, 7500 AE Enschede, The Netherlands
\aff{2}Departmento de Ingenier\'{i}a T\'{e}rmica y de Fluidos, Gregorio Mill\'{a}n Institute for Fluid Dynamics, Nanoscience and Industrial Mathematics, Universidad Carlos III de Madrid, 28911 Leganes, Spain
\aff{3}Canon Production Printing B. V., P.O. Box 101, 5900 MA Venlo, The Netherlands
\aff{4}Department of Chemical and Materials Engineering, University of Alberta, Edmonton T6G1H9, Alberta, Canada
\aff{5}Max Planck Institute for Dynamics and Self-Organization, Am Fassberg 17, 37077 G\"{o}ttingen, Germany
\aff{6}Physical Chemistry and Soft Matter, Wageningen University and Research, Stippeneng 4, 6708 WE Wageningen, The Netherlands}

\newcommand{\wn}{\tilde{w}}
\newcommand{\zn}{\tilde{z}}
\newcommand{\tn}{\tilde{t}}
\newcommand{\Ln}{\tilde{L}}

\newcommand{\tW}{\tilde{w}}
\newcommand{\tT}{\tilde{t}}
\newcommand{\tZ}{\tilde{z}}

\newcommand{\wh}{\hat{w}}

\begin{document}
\maketitle

\begin{abstract}

Evaporation of multi-component liquid mixtures in confined geometries, such as capillaries, is crucial in applications such as microfluidics, two-phase cooling devices, and inkjet printing. Predicting the behaviour of such systems becomes challenging because evaporation triggers complex spatio-temporal changes in the composition of the mixture. These changes in composition, in turn, affect evaporation. In the present work, we study the evaporation of aqueous glycerol solutions contained as a liquid column in a capillary tube. Experiments and \coltext{direct numerical simulations} show three evaporation regimes characterised by \coltext{different temporal} evolutions of the normalised mass transfer rate (or Sherwood number, $Sh$), namely $Sh (\tn) = 1$, $Sh \sim 1/\sqrt{\tn}$, and $Sh \sim \exp\left(-\tn\right)$, where $\tilde{t}$ is a normalised time. We present a simplistic analytical model which shows that the evaporation dynamics can be expressed by the classical relation  $Sh = \exp \left( \tn \right) \mathrm{erfc} \left( \sqrt{\tn}\right)$. For small and medium $\tn$, this expression results in the first and second of the three observed scaling regimes, respectively. This analytical model is formulated in the limit of pure diffusion and when the penetration depth $\delta(t)$ of the diffusion front is much smaller than the length $L(t)$ of the liquid column. When $\delta \approx L$, finite length effects lead to $Sh \sim \exp\left(-\tn\right)$, i.e. the third regime. Finally, we extend our analytical model to incorporate the effect of advection and determine the conditions under which this effect is important. Our results provide fundamental insight into the physics of selective evaporation from a multi-component liquid column.

\end{abstract}


\section{Introduction} \label{sec:intro}

Evaporation of multicomponent volatile liquids into a gaseous phase is ubiquitous in both biological and industrial processes \citep{Erbil2012,Lohse2020,Bourouiba2021,Morris2022,lohse-2022-arfm}. A multicomponent liquid can consist of multiple solvents, surfactants, polymers, colloids, and salts. Evaporation from such systems lead to a plethora of phenomena such as instabilities \citep{Li2020_JFM}, phase separation \citep{tan2016}, \coltext{altering of deposition patterns \citep{palacios-2012-expfluids}}, crystallisation \citep{Shahidzadeh2008}, stratification \citep{hooiveld-2023-jcis}, and evaporation-driven flows \citep{Deegan1997,Hu2006}. In addition to the composition of the liquid, geometrical confinement also affects evaporation significantly. The geometrical confinement can be a droplet \citep{Picknett1977,Lohse2020}, a liquid film \citep{Okazaki1974}, a porous membrane \citep{Prat2002}, a shallow pit \citep{Ambrosio_2021}, or a capillary \citep{Chauvet2009}.

Understanding the evaporation of liquids from capillaries is crucial for applications such as microfluidics \citep{Zimmermann2005, Salmon2022}, inkjet printing \citep{lohse-2022-arfm, rump2022selective}, heat pipes \citep{Chen2016}, chromatography \citep{Kamp2005}, and the measurement of material properties \citep{Roger2018, Nguyen2022, OmerPNAS2022}. Capillaries are also \coltext{considered to be idealised systems} for modelling porous structures \citep{Yiotis2007, Chauvet2009, Chen2022, castell-2023-langmuir}, the transport of water across skin \citep{Sparr2000,Roger2016}, and film drying \citep{Guerrier1998, Salmon2017}.

Capillary evaporation is mainly determined by the behaviour of the volatile liquid meniscus. There have been several experimental and numerical studies to determine the evaporation from a liquid meniscus. These studies describe the evaporation rate \citep{Wayner1971}, heat transfer coefficients \citep{Wayner1976,Park2003,Dhavaleswarapu2012,Zhou2018}, shape of the meniscus \citep{Potash1972,Swanson1992a}, capillary flows that replenish the evaporated liquid \citep{Potash1972,Ransohoff1988,Park2003}, and additional flows that might be driven by evaporation-induced surface tension gradients \citep{Schmidt1992,Buffone2003,Dhavaleswarapu2007,Cecere2014} or buoyancy \citep{Dhavaleswarapu2007,Buffone2019}. 

\begin{figure}
\centering
\includegraphics[width=\textwidth]{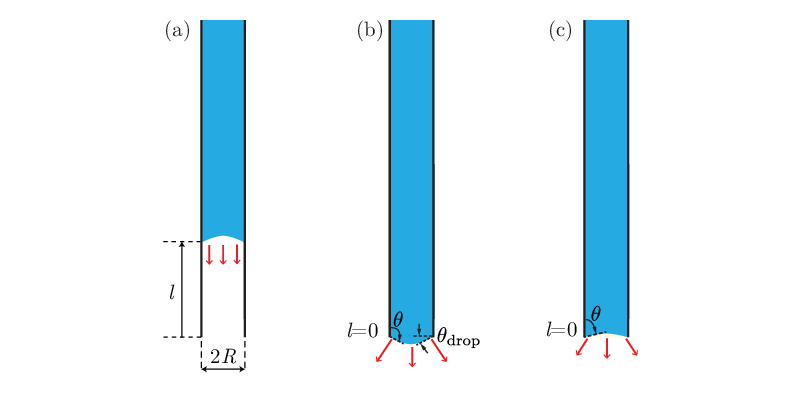}
\caption{ Schematic of the different configurations of evaporation from a capillary: (a) first class of problems, where the liquid meniscus is away from the mouth of the capillary $(l>>0)$; (b, c) second class of problems where the liquid meniscus is close to the mouth of the capillary $(l \approx 0)$, with (b) $\theta>90^\circ$ and (c) $\theta < 90^\circ$. Red arrows represent the \coltext{evaporative} flux.}
\label{fig:problem_types}
\end{figure}

Broadly speaking, the evaporation of a single-component liquid from a capillary can be divided into two main classes depending on the location of the liquid-air meniscus with respect to the open end of the capillary (henceforth referred to as its `mouth', see figure~\ref{fig:problem_types}a). In the first class of problems, the liquid-air interface is far away from the mouth of the capillary. In such a configuration, the evaporation rate is limited by the transport of vapour from the liquid-air interface to the mouth of the capillary tube and it varies approximately as the inverse square root of time \citep{Stefan1873, Stefan1889}. 

In the second class of problems, the liquid meniscus is at (or relatively close to) the mouth of the capillary \citep{Chauvet2009, Gazzola2009}. Within this second class of problems, if the contact angle inside the liquid between the liquid-gas interface and the capillary wall is $\theta \geq 90^\circ$, one can immediately realise the resemblance to a sessile droplet (figure \ref{fig:problem_types}b). In such a case, one can use the Popov model \citep{Popov2005,Junhui2019} to predict the evaporation rate. For $130^\circ > \theta \geq 90^\circ$ (which is equivalent to a contact angle $40^\circ > \theta_{\text{drop}} \geq 0^\circ$ in the case of a sessile droplet), the evaporation rates will be practically independent of the contact angle \citep{Brutin2011_langmuir}, and mainly depend on the base radius, ambient humidity, and the properties of the liquid.

However, when $\theta \leq 90^\circ$, the droplet model for evaporation is no longer applicable (figure~\ref{fig:problem_types}c). The evaporation under such conditions shows two distinct regimes -- a ``constant rate period" and a ``falling rate period" \citep{Chauvet2009, Keita2014, Keita2016} -- similar to that observed \coltext{during} the drying of a porous medium \citep{Coussot2000}. During the constant rate period, evaporation is still mainly determined by the ambient conditions. To replenish the liquid lost by evaporation, upstream liquid is driven by capillary pressure to the mouth of the capillary through thin liquid films \citep{Ransohoff1988,Chauvet2010}. However, even during the so-called constant rate period, the evaporation rate decreases slightly \citep{Coussot2000, Chauvet2009}. For square capillaries, this slight decrease is due to the thinning of liquid films at the mouth of the capillary \citep{Chauvet2009}. As evaporation proceeds further, the depinning of the liquid films from the mouth of the capillary leads to the falling rate period \citep{Chauvet2009}. When the liquid meniscus is sufficiently far away from the mouth of the capillary, the situation reverts to the first class of problems (figure \ref{fig:problem_types}a).

In addition to the studies on evaporation of single component liquids from capillaries, there have also been several studies on evaporation \coltext{in} multi-component systems, such as evaporation of binary liquid mixtures \citep{Duursma2008, Cecere2014, Salmon2017, Zhou2018}, surfactant solutions \citep{He2015, Roger2016, Roger2018}, salt solutions \citep{Camassel2005, Naillon2015}, and colloidal dispersions \citep{Abkarian2004, Kamp2005, Sarkar2009, Wang2020a, Roger2021} in capillaries. In all these scenarios, evaporation leads to spatio-temporal variations in the composition of the mixture. Nevertheless, it is generally possible to model the capillary evaporation of a multi-component system as a one-dimensional transport problem. Unsurprisingly, but perhaps interesting to note is that in the absence of instabilities \citep{deGennes2001}, the one-dimensional model of evaporation of polymeric liquid films \citep{Okazaki1974,Guerrier1998,Saure1998,Okuzono2006} is mathematically-equivalent to the evaporation of binary solutions from capillaries \citep{Salmon2017}.

In multi-component systems, especially binary systems, the evaporation rate can also show a constant rate period followed by a falling rate period \citep{Okazaki1974,Salmon2017,Poon2022}, similar to pure liquids. However, in multi-component systems, the different evaporation regimes are additionally determined by changes in the composition of the mixture. Hence, a complete evaporation model must include the spatio-temporal variations in the composition and properties of the mixture. Recently, \citet{Salmon2017} showed how a steep decrease in thermodynamic activity and diffusion coefficient of water at high solute concentrations can lead to evaporation being almost independent of ambient humidity for certain molecularly complex fluids. These authors modelled the variable diffusion coefficient as a piece-wise constant function, but bypassed the necessity of using an analytical expression for the thermodynamic activity of water. Moreover, \citet{Salmon2017} considered the parameter range where the medium can be approximated as semi-infinite. 

In this work, we study the evaporation of binary liquids in capillaries with experiments, direct numerical simulations, and analytical modelling. We perform experiments for the evaporation of glycerol-water mixtures in a cylindrical capillary tube under controlled humidity conditions. Our direct numerical simulations show excellent agreement with the experiments. Further, to unravel the physics of the evaporation dynamics, we develop a one-dimensional analytical model. We introduce a linear approximation for the thermodynamic activity of water as a function of its weight fraction. We also identify the condition when the semi-infinite assumption breaks down, and accordingly take it into account in our modelling. Finally, we discuss the explicit role of the advective mass transport compared to the diffusive mass transfer for our particular system. We show that our model accurately predicts the relevant scaling laws observed in the experiments and the numerical simulations. 

The paper is organised as follows: in \S~\ref{sec:exp} we describe the experimental setup and observations. The governing equations of our system and the numerical method are described in \S~\ref{sec:problemForm}. In \S~\ref{sec:modellingStart}, we provide three simplified analytical models of the problem, each with an added level of complexity over its predecessor, and compare their predictions with the results obtained from the experiments and the simulations. The manuscript culminates with a summary of the results and an outlook in \S~\ref{sec:conclusion}. 

\section{Experiments} \label{sec:exp}

\begin{figure}
\centering
\includegraphics[width=\textwidth]{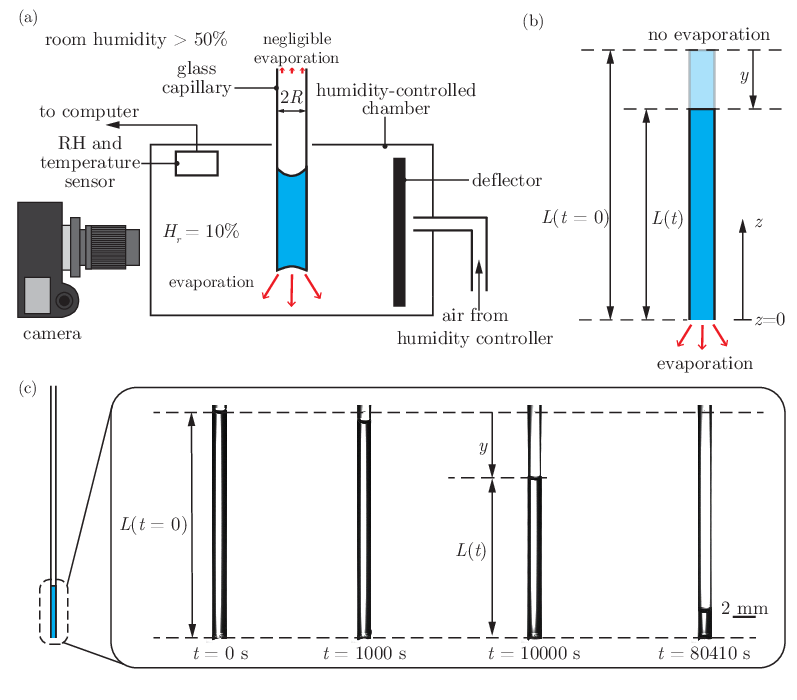}
\caption{ (a) Schematic of the experimental setup. (b) Geometry used for numerical simulation and analytical modelling. (c) Left: schematic of the relative length of the liquid column with respect to the length of the capillary; right: time-lapsed experimental snapshots of water-glycerol mixtures in the capillaries for \coltext{an} initial weight fraction of water $w_{i}$ = 0.9. The top interface of the liquid mixture keeps moving downwards due to the evaporation of water from the bottom of the capillary. Red arrows \coltext{in (a) and (b) represent the evaporative flux.}}
\label{fig:setup}
\end{figure}

\subsection{Experimental setup}

Aqueous solutions of different mass fractions of glycerol (Sigma-Aldrich) were used as the probe liquids in the present experiments. The reader is referred to \citet{sen-draft} for details of the preparation procedure of the liquids and their characterisation. The use of glycerol has the following advantages. Firstly, since glycerol has very low vapour pressure, it is practically non-volatile under the current experimental conditions. Thus we only need to account for the evaporation of water. Further, since the more volatile liquid (water) has higher surface tension, evaporation of water should not lead to any flow instabilities close to the interface \citep{Diddens2017a}. Such instabilities occur whenever the mass transfer (evaporation or condensation) leads to an increase in surface tension, such as for evaporating water-ethanol mixtures \citep{Christy2011, bennacer_sefiane_2014, diddens_jfm_2017, RicardoJFM2021} or condensation of water onto a water-glycerol droplet \citep{Stone_Glycerol2016,Diddens2017a}. A more detailed explanation of such instabilities can be found in \citet{Diddens2017a}.

In the present experiments, we studied the evaporation dynamics of aqueous solutions of glycerol in a thin cylindrical capillary tube (figure \ref{fig:setup}; inner diameter = \SI{1}{\milli\meter}, outer diameter = \SI{1.2}{\milli\meter}; length = \SI{100}{\milli\meter}, Round Boro Tubing, CM Scientific). The liquid column inside the capillary had an  initial height of 19 $\pm$ 2 \si{\milli\meter}. The initial weight fraction of water, $w_i$, in the water-glycerol mixture was varied as 0.2, 0.6, 0.9, and 1.0, to cover a wide range of initial compositions. The lower end of the capillary was placed inside an in-house developed, optically-transparent, humidity-controlled chamber at room temperature. The humidity and temperature inside the chamber were monitored using a temperature-humidity sensor (HIH6121, Honeywell). The relative humidity in the chamber was maintained at $H_r = 10 \pm 5 \%$. The upper end of the capillary tube was exposed to room humidity ($>50\%$). The evaporation or condensation of water at the upper meniscus was negligible compared to the evaporation from the bottom. This is because of the \coltext{relatively} large distance between the liquid's upper meniscus and the capillary's upper end (see appendix  \ref{section:appA} for a detailed discussion).

The contact line of the lower meniscus remained pinned at the lower mouth of the capillary. Thus, the loss of water by evaporation from the lower mouth of the capillary leads to a decrease in the length $L$ of the liquid column (figure \ref{fig:setup}c). To study this evaporation process quantitatively, time-lapsed images of the liquid column were captured using a DSLR camera (D750, Nikon) equipped with either a long-distance microscope (Navitar 12$\times$) or a macro lens (50 mm DG Macro D, Sigma), while the capillary tube was back-illuminated with a cold LED light source (Thorlabs). For pure water, the velocity $v_y$ of the upper interface, 
\begin{equation}
    v_y = \frac{\mathrm{d}y}{\mathrm{d}t}= \frac{1}{\rho_w} \frac{\mathrm{d}M^{\prime \prime}}{\mathrm{d}t} ,
\label{eqn:vy_pure}
\end{equation}
is a direct measure of the evaporation rate $\mathrm{d}M^{\prime \prime}/\mathrm{d}t$ of water, where $y$ is the displacement of the top interface, $M^{\prime \prime}$ the mass per unit area, and $\rho_w$ the density of water. 

At a later time, the contact line of the lower liquid meniscus eventually depins. At this point, we stop the measurements because $v_y$ is thereafter no longer a correct measure of the evaporation rate. Additionally, as the lower interface moves inwards into the capillary tube after depinning, the evaporation boundary condition at the lower interface also changes (see \S~\ref{sec:theoretical-model} for details of the boundary conditions).

\subsection{Experimental results}

\begin{figure}
\centering
\includegraphics[width=\textwidth]{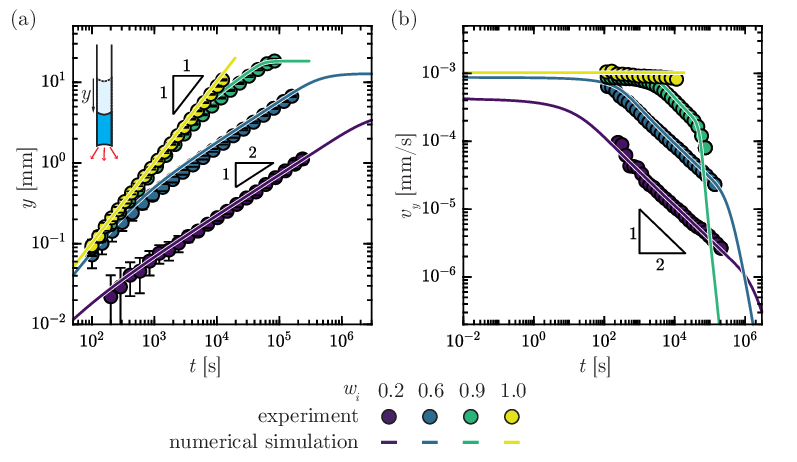}
\caption{ (a) Displacement $y$ and (b) velocity $v_y$ of the top interface of the aqueous solutions of glycerol observed in experiments (discrete datapoints) and numerical simulations (continuous lines), for initial weight fractions of water $w_{i}$ = 0.2, 0.6, 0.9, and 1.0.}
\label{fig:ytvt_Plots}
\end{figure}

The discrete datapoints in figure \ref{fig:ytvt_Plots}a denote the experimentally-obtained vertical displacement $y$ of the upper interface with time $t$ for different initial weight fractions of water, $w_i$. Only every fifth datapoint is plotted in figure \ref{fig:ytvt_Plots}a to avoid overcrowding of the plot. The markers represent the mean of at least three independent experimental realizations, while the errorbars (\coltext{denoting} $\pm$ one standard deviation) reflect the uncertainty due to experimental variabilities and the resolution of the imaging. The velocity $v_y = \mathrm{d}y/ \mathrm{d}t$ of the top interface, calculated from $y$, is plotted as discrete datapoints in figure \ref{fig:ytvt_Plots}b.

The motion of the top interface obviously depends on the initial composition of the liquid mixture (figure \ref{fig:ytvt_Plots} and Supplementary Video V1). For pure water ($w_i=1$), the interface moves at an almost constant velocity ($y=v_y\,t$ and $\coltext{v_y(t)} \approx 10^{-3}$ mm/s, see figure \ref{fig:ytvt_Plots}). For $w_i = 0.2$ and $0.6$, the experiments suggest $y \sim t^{1/2}$ and $v_y \sim t^{-1/2}$ scaling relations. However, for $w_i=0.9$, the experiments show three different regimes: $\coltext{v_y(t)} \approx 0.8 \times 10^{-3}$~mm/s, similar to $w_i = 1$, in the first $\approx 800$ s; $v_y \sim t^{-1/2}$, similar to $w_i=0.2$ and $0.6$, at intermediate times ($t \lessapprox 5 \times 10^4$ s); and a decrease in velocity (which is steeper than $v_y \sim t^{-1/2}$) at long times.

The addition of glycerol primarily reduces the local concentration of water at the lower interface, which in turn leads to a reduced evaporation rate and a decrease in the velocity of the upper interface. Thus, we get the so-called constant rate period and the so-called falling rate period as described in literature \citep{Salmon2017}. Our experimental results, however, raise \coltext{two} important questions. Firstly, why does the constant rate period not appear for $w_i=0.2$ and $0.6$ in the experiments? Secondly, why does the falling rate period show two sub-regimes for $w_i=0.9$? To answer these questions, and to understand how the evaporation dynamics of a water-glycerol mixture in a capillary change with time and the initial composition of the mixture, we develop a theoretical model in the following section.

\section{Problem formulation} \label{sec:problemForm}

\subsection{Theoretical model} \label{sec:theoretical-model}

We model the column of water-glycerol mixture as a one-dimensional isothermal system with length $L(t)$ (figure \ref{fig:setup}b). The lower end of the capillary is located at $z=0$. We can express the \coltext{liquid} composition using the weight fraction of water $w(z,t)$; the weight fraction of glycerol then simply is $1- w(z,t)$. The spatio-temporal variations in the concentration of water in the liquid column can be determined by solving the one-dimensional continuity and advection-diffusion equations:
\begin{equation}
    \frac{\p \rho}{\p t} + \frac{\p}{\p z} (\rho \, u) = 0
    \label{eqn:continuity}
\end{equation}
and
\begin{equation}
\frac{\p}{\p t} (\rho \, w) + u \, \frac{\p}{\p z} (\rho \, w) =   
\frac{\p}{\p z} \, \left( \rho \, D \, \frac{\p w}{\p z} \right) ,
    \label{eqn:FullMassTransport}
\end{equation}  
where $\rho (w)$ is the local density of the mixture, $D(w)$ the diffusion coefficient of the water-glycerol mixture, and $u$ the fluid velocity along the $z$-direction.

Initially (at $t=0$), the composition in the liquid column is uniform and equal to $w_i$. Since the evaporation of water from the upper interface is negligible (see appendix \ref{section:appA}), we model the upper interface to be non-evaporative. The lower interface of the liquid column is exposed to the ambient at a constant relative humidity of $H_{r}$ and loses water by evaporation (per unit area) at a rate of $\mathrm{d}M^{\prime \prime}/\mathrm{d}t$. The water lost due to evaporation is replenished by the diffusive and advective transport of water from the bulk liquid inside the capillary. These considerations lead to \coltext{one initial and two boundary conditions as follows}:

\begin{equation}
 \begin{array}{cl}
\displaystyle w = w_i 
    & \quad \mbox{at\ }\quad t=0\, ,\\[8pt]
\displaystyle\frac{\p w}{\p z} = 0 
    & \quad \mbox{at\ }\quad z = L(t) \, , \\[8pt]
\displaystyle -\rho \, u \, w + \rho \, D \, \frac{\p w}{\p z} = \frac{\mathrm{d}M^{\prime \prime}}{\mathrm{d}t}
    & \quad \mbox{at\ }\quad z=0 \, .
 \end{array}
  \label{eqn:bc_main}
\end{equation}

The evaporation of water from the bottom interface can be approximated using a quasi-steady diffusion-limited model of evaporation of a pinned sessile droplet having zero contact angle \citep{Popov2005,stauber2014}: 
\begin{equation}
    \frac{\mathrm{d}M^{\prime \prime}}{\mathrm{d}t} = \pi \, D_{v,a} \, R \, (c_{w,s}-c_{w,\infty}) \, f(\theta_{\text{drop}}) \, \frac{1}{\pi \, R^2} ,
    \label{eqn:Popov}
\end{equation}
where $D_{v,a}$ is the diffusion coefficient of water vapour in air, $R$ the inner radius of the capillary tube, $c_{w,s}$ the concentration (vapour mass per volume) of water vapour in the air at the lower interface, $c_{w,\infty}=H_r c_{w,s}^o$ the concentration of water vapour in the ambient air (far away from the capillary), $c_{w,s}^o$ the saturation concentration of water vapour at the surface of pure water, $\theta_{\text{drop}}$ the angle between the horizontal plane and the liquid-air interface at the contact line, and $f(\theta_{\text{drop}})$ a known function of $\theta_{\text{drop}}$. For $\theta_{\text{drop}}=0$ as in our model here, $f(\theta_{\text{drop}}) = 4/\pi$  \citep{Popov2005, stauber2014}.
Thus, 
\begin{equation}
    \frac{\mathrm{d}M^{\prime \prime}}{\mathrm{d}t} =\frac{4 D_{v,a}}{\pi \, R} \, (c_{w,s}-c_{w,\infty}) \, .
    \label{eqn:finalPopov}
\end{equation}

\noindent $c_{w,s}$ depends on the composition of the liquid mixture close to the interface (at $z=0^+$) and is given by Raoult's Law as
\begin{equation}
    c_{w,s} = a \, c_{w,s}^o = x_{o} \, \psi_{o} \, c_{w,s}^o \, ,      
    \label{eqn:cws}
\end{equation}
where $a(w)$ is the thermodynamic activity of water, $x_{o}$ the mole fraction of water in the liquid at $z=0^+$, and $\psi_o(x_o)$ the activity coefficient of water corresponding to $x_o$. As long as $c_{w,s}>c_{w,\infty}$, water will evaporate from the lower interface. 

\subsection{Numerical solution} \label{sec:numerics}

We numerically solve \coltext{the equations pertaining to} the aforementioned one-dimensional theoretical model using Finite Element Method simulations with an initial length of $L=\SI{20}{\milli\meter}$ to obtain the evaporation rates and the spatio-temporal distribution of the concentration of water $w$ in the capillary. To that end, a line mesh initially consisting of 100 second order Lagrangian elements is created to cover the initial length $L$. The motion of the top interface is realized by moving the mesh nodes along with the top interface, i.e. by an arbitrary Lagrangian-Eulerian method (ALE) with a Laplace-smoothed mesh. The mesh displacement at the top interface $z=L(t)$, i.e. $\dot L(t)=u(L,t)$, is enforced by a Lagrange multiplier acting on the node position at the top. Likewise, the evaporation at the bottom is considered, but here the Lagrange multiplier is acting on the velocity $u$ at $z=0$, whereas the bottom node remains fixed at $z=0$.

The implementation of \eqref{eqn:FullMassTransport} and \eqref{eqn:continuity} along with the boundary conditions \eqref{eqn:bc_main} and \eqref{eqn:finalPopov} is achieved by the conventional weak formulations of advection-diffusion equations, including the ALE corrections for the time derivatives. \textsl{A posteriori} spatial adaptivity based on the jumps in the slopes of $w$ across the elements is considered. Also, \textsl{a posteriori} temporal adaptivity is considered by calculating the difference between the freshly calculated value of each field at each point and its prediction. For the prediction, values from the previous time steps are extrapolated to the current time. If the difference is large, this means that the system changes excessively during a time step. In that case, the current time step is rejected and calculations are made again with a smaller time step. The implementation has been done by the finite element library \textsc{oomph-lib} by \citet{Heil2006}, which monolithically solves the coupled equations with a backward differentiation formula of second order for the temporal integration.

The variation of the diffusion coefficient $D$ as a function of the local composition $w$ is considered in the simulations based on the experimental data of \citet{DErrico2004}, while the mass density $\rho$ was fitted according to the data of \citet{Takamura2012}. The activity coefficient of water was calculated by AIOMFAC \citep{Zuend2011}.

The results of the direct numerical simulations are shown by the continuous lines in figure~\ref{fig:ytvt_Plots}. Excellent agreement between the experiments and the numerical simulations is observed. In particular, figure \ref{fig:ytvt_Plots}b shows that the simulations can reproduce the experimentally observed $v_y \sim t^{-1/2}$ scaling for $w_i=0.2$ and $0.6$, and all the three velocity scalings for $w_i=0.9$. Interestingly, the simulations also show that for very early times, $v_y$ is almost constant for $w_i=0.2$ and $0.6$ as well. However, we cannot access these time scales in experiments due to limitations arising from the lack of spatio-temporal resolution. Overall, the quantitative match between experiments and numerical simulations show that our theoretical model incorporates all the relevant physics of the problem. In figure \ref{fig:axialDistribution} of appendix \ref{section:appAxialConc}, we also show the spatial variation in the axial \coltext{concentration} profiles for the different initial conentrations $w_i$. In the next section, we will use additional simplifying assumptions to formulate a simplistic model which captures the essential physics of the system and recovers the various evaporation regimes.

\section{Analytical model} \label{sec:modellingStart}
 
We present here a simplified description of the problem with the purpose of elucidating the physical mechanisms behind the different regimes observed in the experiments and the direct numerical simulations. 
We introduce some assumptions that will allow us to treat the resulting problem analytically. As we will show below, despite these simplifications, the quantitative comparison between the model and the experiments and simulations is reasonably good. 
Our one-dimensional analytical model relies on the following assumptions:
 
\noindent (i) Constant properties: we assume that the properties of the water-glycerol mixture, namely, density $\rho$ and diffusion coefficient $D$, are constant and equal to the values corresponding to the initial composition. These properties can be obtained from figure \ref{fig:materialProp} in appendix \ref{section:app_property} by setting $w = w_i$. Setting density as constant in the continuity equation \eqref{eqn:continuity} yields that $u$ is independent of $z$ and only depends on $t$. Hence, 
\begin{equation}
    u(z,t) = - v_y(t) .
    \label{eqn:const_u}
\end{equation}

\noindent (ii) Linearisation of the water vapour concentration difference: the concentration of water vapour at the liquid-gas interface depends on the concentration of water at $z=0^+$  \eqref{eqn:cws}. To solve the model analytically, we linearise the expression in \eqref{eqn:finalPopov} for the difference in concentration of the water vapour between $c_{w,s}(w_i)$ and $c_{w,s}(w_\mathrm{eq})$ in terms of $w$ as 
\begin{equation}
    c_{w,s} - c_{w,\infty} = c_{w,s}^0 \, \frac{x_i \, \psi_i - H_r}{w_i - w_\mathrm{eq}} \,  (w \vert_{z=0} - w_\mathrm{eq}) ,\label{eqn:deltaCs_linear}
\end{equation} 
where $x_i$, $\psi_i$, and $w_i$ are the initial mole fraction, activity coefficient, and weight fraction of water in the liquid mixture, respectively, and $w_\mathrm{eq}$ the weight fraction of water at equilibrium, i.e. when $c_{w,s}$  becomes equal to $c_{w,\infty}$ and evaporation stops. In \eqref{eqn:deltaCs_linear}, we have effectively linearised $\left( c_{w,s}-c_{w,\infty} \right)$ in terms of $w$, between the initial ($w=w_i$) and final ($w=w_\mathrm{eq}$) concentrations of water (see detailed derivation in appendix \ref{section:app_Linearisation}). Combining \eqref{eqn:finalPopov} and \eqref{eqn:deltaCs_linear}, we get
\begin{equation}
    \frac{\mathrm{d}M^{\prime \prime}}{\mathrm{d}t} = h^{\ast} \,  (w \vert_{z=0} - w_{eq}) ,
    \label{eqn:dmdt_linear_shortened}
\end{equation}
where $h^{\ast}$, defined as
\begin{equation}
    \coltext{h^{\ast} = \frac{4D_{v,a} \, c_{w,s}^0 }{\pi \, R \, \Delta w_i} \, (x_i \, \psi_i - H_r) ,}
\end{equation}
is a modified mass transfer coefficient \coltext{and $\Delta w_i = w_i - w_\mathrm{eq}$}. We put an asterisk in $h^{\ast}$ to denote that its units (kg/m$^{2}$s) are different from those of the conventional mass transfer coefficient $h$ \citep{Incropera2007}, which is related to $h^*$ as $h=h^*/\rho$ (unit of m/s). 

\noindent (iii) Velocity of meniscus: assuming that the volume of the glycerol-water mixture is the sum of the glycerol and water volumes, the velocity at which the length of the liquid column recedes is given by
\begin{equation}
    v_y = -\frac{\mathrm{d}}{\mathrm{d}t}\int_0^L \frac{\rho w}{\rho_w}\,\mathrm{d}z,
\end{equation}
i.e. the negative of the time derivative of the volume occupied by the water, since the glycerol volume is constant. \coltext{Since $\mathrm{d}M^{\prime \prime}/\mathrm{d}t$ denotes} the rate at which the water mass per unit cross-section of capillary is lost, \coltext{we get}
\begin{equation}
    v_y = \frac{1}{\rho_w} \frac{\mathrm{d}M^{\prime \prime}}{\mathrm{d}t} \, .
    \label{eqn:vy_mix}
\end{equation}

\noindent \coltext{Note} that this equation is identical to the exact expression \eqref{eqn:vy_pure} obtained for the case where the liquid just contains water. This is a direct consequence of the nearly ideal character of the glycerol-water mixtures.

We define a diffusive length scale $l_D$ (as also done by \citet{Salmon2017}) based on the mass transfer coefficient by considering a balance between evaporation and diffusion at the lower interface ($D \Delta w / l_D \sim h \Delta w$; \eqref{eqn:bc_main} and \eqref{eqn:dmdt_linear_shortened}) yielding

\begin{equation}
    l_D = \frac{D}{h}. 
\end{equation}

Based on the aforementioned considerations, we non-dimensionalise the variables as follows:
\begin{equation}
    \coltext{\wn = \frac{w_i - w}{w_i - w_\mathrm{eq}} = \frac{w_i-w}{\Delta w_i}}, \quad
    \tn = \frac{t}{l_D^2/D} = \frac{h^2 \,  t}{D}, \quad
    \zn = \frac{z}{l_D} = \frac{h \, z}{ D}, \quad
    \Ln = \frac{L}{l_D} = \frac{h \, L}{D}. \quad
     \label{eq:normdef}
\end{equation}

\noindent Further, since the velocity of the interface is in fact a proxy for the mass transfer (evaporation) rate of water, \coltext{as follows from mass conservation, we} can describe the system in terms of the Sherwood number, $Sh$. \coltext{This parameter denotes the} non-dimensional velocity or non-dimensional mass transfer rate:
\begin{equation} 
    Sh = \frac{\rho_w \, v_{y}}{h^{\ast} \coltext{\Delta w_i}} . \label{eqn:ShDefine}
\end{equation}
\coltext{Substituting \eqref{eqn:const_u}-\eqref{eqn:ShDefine} into the governing differential equations \eqref{eqn:continuity} and \eqref{eqn:FullMassTransport} of mass transport in the capillary, the initial and boundary conditions \eqref{eqn:bc_main}, and the equation governing evaporation of water into air \eqref{eqn:finalPopov}}, we get the following system of equations, and initial and boundary conditions:
\begin{equation}
    \frac{\p \wn}{\p \tn} - \coltext{\frac{\rho \Delta w_i Sh}{\rho_w}} \, \frac{\p  \wn}{\p \zn} = \frac{\p^2 \wn}{\p \zn^2} ,
    \label{eqn:neq_c0}
\end{equation}

\begin{equation}
 \begin{array}{cl}
\displaystyle\wn = 0 
    &\quad \mbox{at\ }\quad \tn=0,\\[8pt]
\displaystyle\frac{\p \wn}{\p \zn} = 0 
    &\quad \mbox{at\ }\quad \zn = \Ln, \\[8pt]
\coltext{\displaystyle \frac{\p \wn}{\p \zn} - \left(1 - \frac{\rho \Delta w_i Sh}{\rho_w}\right) \, \wn + \left( 1 - \frac{\rho w_i Sh}{\rho_w}\right) = 0}
    &\quad \mbox{at\ }\quad \zn=0 \, ,
 \end{array}
  \label{eqn:nbc_c0}
\end{equation}

\begin{equation}
    \coltext{Sh = 1 - \tW(\tZ=0, \tT) \, . }
    \label{eqn:Sh_w_relation}
\end{equation}

Finally, for the case of pure water, we do not need to solve the model for the binary mixture, since there is no change in the concentration. The velocity of the interface can then be directly obtained by substituting $w_i=1$ in \eqref{eqn:dmdt_linear_shortened}:
\begin{equation}
    v_y = \frac{4D_{v,a}}{\pi \, R \, \rho_w} \,  c_{w,s}^0 \, \left(1 - H_r \right),
    \label{eqn:vt_water}
\end{equation}
\noindent which corresponds to $Sh=1$.

\subsection{Semi-infinite transient diffusion model} \label{sec:model1}

\begin{figure}
\centering
\includegraphics[width=\textwidth]{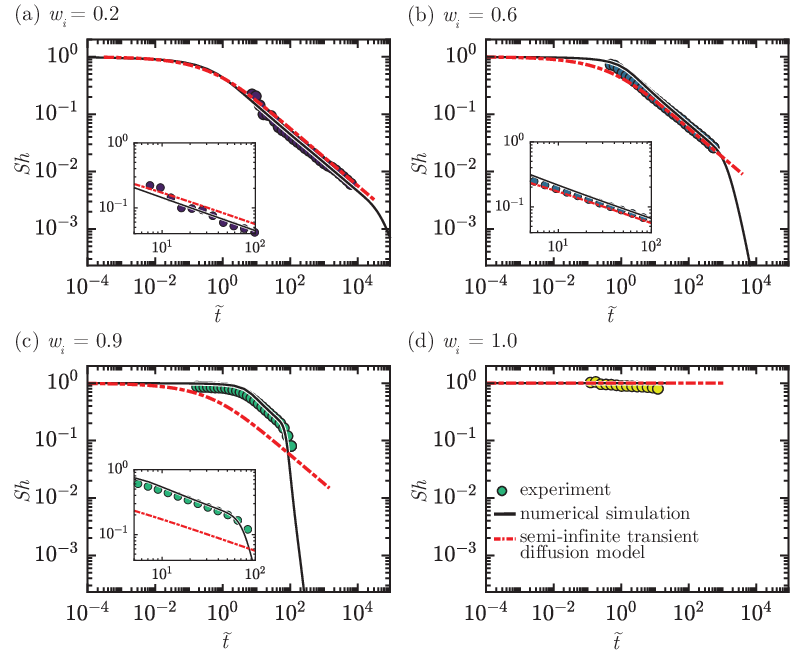}
\caption{ (a) Comparison of the normalised evaporation rates (Sherwood number, $Sh$) against normalised time ($\tn$) obtained from the experiments (discrete datapoints) and the numerical simulations (continuous lines) with theoretical values (dashed lines) obtained using the semi-infinite transient diffusion model \eqref{eqn:ShBimodel1}, for different initial weight fractions of water $w_i$. Notice that in the case of $w_i = 1.0$ (panel (d)), the theoretical curve corresponds to $Sh = 1$, and not that given by \eqref{eqn:ShBimodel1}. Note also that all the theoretical curves for $w_i<1.0$ are \coltext{the same, as \eqref{eqn:ShBimodel1} does not take into account $w_i$. Insets in panels (a), (b), and (c) highlight the comparisons between the experiments, the numerical simulations, and the simplified analytical modelling. }}
\label{fig:SemiInfiniteModel}
\end{figure}

As the zeroth order simplification, we assume that advection is negligibly small. Further, we approximate the liquid column as a semi-infinite medium. Thus, the governing equation and the initial and boundary conditions (\eqref{eqn:neq_c0} and \eqref{eqn:nbc_c0}) can be reduced to:
\begin{equation}
    \frac{\p \wn}{\p \tn} = \frac{\p^2 \wn}{\p \zn^2} ,
    \label{eqn:neq_c1}
\end{equation}
with
\begin{equation}
 \begin{array}{cl}
\displaystyle\wn = 0 
    &\quad \mbox{at\ }\quad \tn=0,\\[8pt]
\displaystyle\frac{\p \wn}{\p \zn} = 0 
    &\quad \mbox{at\ }\quad \zn \rightarrow \infty, \\[8pt]
\displaystyle \frac{\p \wn}{\p \zn} = \wn - 1 
    &\quad \mbox{at\ }\quad \zn=0 .
 \end{array}
  \label{eqn:nbc_c1}
\end{equation}
This is a classical transient diffusion problem with mixed boundary conditions, whose solution is given by \citep{Incropera2007}:
\begin{equation}
    \wn \left( \zn,\tn \right) = \mathrm{erfc}\left( \frac{\zn}{2\sqrt{\tn}} \right) -\exp \left(\tn + \zn\right)\mathrm{erfc}\left(\frac{\zn}{2\sqrt{\tn}} + \sqrt{\tn}\right) .
    \label{eqn:wn_sol1}
\end{equation}
The velocity of the top interface can be now evaluated from the evaporation rate of water \coltext{using \eqref{eqn:ShDefine}, \eqref{eqn:Sh_w_relation}, and \eqref{eqn:wn_sol1}} to yield
\begin{equation}
    v_y = \frac{h^{\ast} \coltext{\Delta w_i}}{\rho_w} \, \exp(\tn) \, \mathrm{erfc}\left(\sqrt{\tn}\right) , 
    \label{eqn:vtmodel1}
\end{equation}
or 
\begin{equation}
    Sh = \exp \left( \tn \right) \, \mathrm{erfc}\left( \sqrt{\tn}\right) .
    \label{eqn:ShBimodel1}
\end{equation}

\noindent For $\tn \rightarrow 0$, $Sh \rightarrow 1$, whereas for $\tn \gg 1$, $Sh \approx 1/\sqrt{\pi \tn}$.

The predictions of the semi-infinite transient diffusion model \eqref{eqn:ShBimodel1} are compared in figure \ref{fig:SemiInfiniteModel} with the experimental measurements (discrete datapoints) and the numerical simulations (continuous lines). It can be observed that \eqref{eqn:ShBimodel1} correctly predicts the early time limit of $Sh=1$ for all cases, and $Sh \sim 1/\sqrt{\tn}$ at intermediate times for $w_i=0.2$ and $0.6$. \coltext{Moreover}, figure \ref{fig:SemiInfiniteModel} also shows that the predicted value of $Sh$ from the model agrees reasonably well with the experiments and the simulations for $w_i=0.2$ and $0.6$. However, \eqref{eqn:ShBimodel1} severely underpredicts $Sh$ for $w_i=0.9$ until $\tn \approx 80$. \coltext{Furthermore}, \eqref{eqn:ShBimodel1} also fails to capture the steep decay in $Sh$ seen in the simulations for $w_i=0.6$ and $w_i=0.9$ at long times (figures \ref{fig:SemiInfiniteModel}b \coltext{and} c). This calls for a careful re-examination of the assumptions made in the model. 

\subsection{\coltext{Semi-infinite transient diffusion model with advection -- asymptotic solution for $\tT \gg 1$ and approximate solution for all times}} \label{sec:model3}

\begin{figure}
\centering
\includegraphics[width=\textwidth]{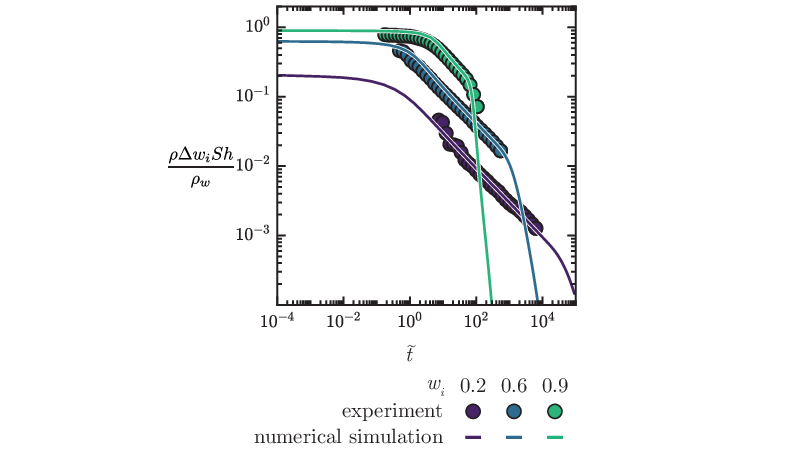}
\caption{\coltext{Plots showing the variation of the Peclet number $\rho \Delta w_i Sh/ \rho_w$ with the normalised time $\tn$ for three different initial water concentrations $w_{i} = 0.2$, 0.6, and 0.9. }}
\label{fig:PecletNumber}
\end{figure}

\coltext{In order to improve the agreement of our model with the experiments and full numerical simulations, we turn our attention towards the effect of advection. Indeed, advection is expected to become important at short times and for values of $w_i$ close to unity. To make this clear, we define the Peclet number of the problem as the coefficient of the advection term in \eqref{eqn:neq_c0}, namely \coltext{$\Delta w_i \, Sh \, \rho / \rho_w = \rho \, v_y / h^{\ast} = (w \vert_{z=0} - w_\mathrm{eq}) \rho/\rho_w$}. Notice that in our problem the Peclet number is nothing more than the Sherwood number with a coefficient that modulates the importance of the initial water concentration. However, we find it useful to work with this quantity in order to evaluate the effect of advection. This Peclet number is plotted in figure \ref{fig:PecletNumber}, where we can see that it becomes of order unity during the first stage of the evaporation process for $w_i \gtrsim 0.6$.}

\coltext{To quantitatively predict the evaporation dynamics including the effect of advection, we propose an improvement over our simplistic analytical model. We will first develop an asymptotic solution, formally valid in the limit $\tT \gg 1$, followed by the presentation of an approximate expression of $Sh$ that converges uniformly to the asymptotic solution for $\tT \gg 1$ and to $Sh \approx 1$ for $\tT \ll 1$.}

\coltext{We start from the problem formulated in \eqref{eqn:neq_c0} and \eqref{eqn:nbc_c0}, with $\tilde{L}\rightarrow\infty$. In order to make the problem tractable, we assume the mixture's density is constant and equal to the initial one, $\rho = \rho(w_i)$. The transport problem \eqref{eqn:neq_c0} with advection becomes,}

\begin{equation}
    \coltext{\frac{\partial\tW}{\partial\tT} - \Delta w_i\,Sh\,\frac{\rho}{\rho_w}\frac{\partial\tW}{\partial\tZ} = \frac{\partial^2\tW}{\partial\tZ^2},}
    \label{eq:simplified_eqn_adv}
\end{equation}
\coltext{with one initial and two boundary conditions \eqref{eqn:nbc_c0}, namely}

\begin{equation}
\begin{array}{cl}
\displaystyle \coltext{\tW = 0} &\quad \mbox{\coltext{at}\ }\quad \coltext{\tT = 0 \, ,}\\
\displaystyle \coltext{\frac{\partial\tW}{\partial\tZ} = 0} &\quad \mbox{\coltext{at\ }}\quad \coltext{\tZ \rightarrow \infty \, ,}\\
\displaystyle \coltext{\frac{\partial\tW}{\partial\tZ} - \left(1 - \Delta w_i \frac{\rho}{\rho_w}\,Sh\right)\tW + 1 - \frac{\rho}{\rho_w}w_i\,Sh = 0} &\quad \mbox{\coltext{at}\ }\quad \coltext{\tZ = 0 \, ,}
\end{array}
    \label{eq:simplified_eqn_adv_bc}
\end{equation}

\noindent \coltext{and with the Sherwood number $Sh$ (the water evaporation rate at the interface) given by} \eqref{eqn:Sh_w_relation}, 

\begin{equation}
    \coltext{Sh = 1 - \tW(\tZ=0, \tT).}
    \label{eq:Sh_w_relation2}
\end{equation}

\coltext{\subsubsection{Asymptotic solution}}

\coltext{We shall now look for solutions at long times. Inspired by both experiments and numerical simulations, we investigate solutions where} 
\begin{equation}
    \coltext{Sh = A\tT^{-\lambda}},
\label{eqn:Sh_A_definition}
\end{equation} 

\noindent \coltext{with $\lambda > 0$. Further, since at long times we expect $\tW$ to approach unity everywhere, we propose the following change of variables:} 

\begin{equation}
    \coltext{W = 1 - \tW - Sh \, .}
    \label{eqn:define_capitalW}
\end{equation} 

\coltext{Thus $W \rightarrow 0$ as $\tW \rightarrow 1$ and $Sh \rightarrow 0$, i.e. at long times. Reformulating the problem in terms of the new variable and neglecting terms $\mathcal{O}(Sh^2)$, \eqref{eq:simplified_eqn_adv} - \eqref{eq:Sh_w_relation2} can be re-written as}
\begin{equation}
    \coltext{\frac{\mathrm{d}Sh}{\mathrm{d}\tT} + \frac{\partial W}{\partial\tT} - \Delta w_i\frac{\rho}{\rho_w}\,Sh\frac{\partial W}{\partial\tZ} = \frac{\partial^2 W}{\partial\tZ^2} \, ,}
    \label{eq:simplified_eqn_adv_W}
\end{equation}

\begin{eqnarray}
    \coltext{W = 1 - Sh} & \;\mathrm{\coltext{at}}\; & \coltext{\tT = 0 \,} ,\label{eq:W_short_t}\\
    \coltext{\frac{\partial W}{\partial\tZ} = 0} & \;\mathrm{\coltext{at}}\; & \coltext{\tZ \rightarrow \infty \, ,}\label{eq:dWdZ0}\\
    \coltext{-\frac{\partial W}{\partial\tZ} + Sh \left(1 + w_{eq}\frac{\rho}{\rho_w}\right)= 0} & \;\mathrm{\coltext{at}}\; & \coltext{\tZ = 0} \, ,\label{eq:boundary_condition_liquid}
\end{eqnarray}

\begin{equation}
    \coltext{W(\tZ = 0, t) = 0 \, .}
    \label{eq:simplified_eqn_W_at_z0}
\end{equation}

\coltext{We seek for self-similar solutions of the type $W(\tZ, \tT) = F(\eta)$, with $\eta = \tZ / \tT^\lambda$, in the limit $\tT \gg 1$. For such a self-similar solution to exist, the exponent $\lambda$ of $\tilde{t}$ in the self-similar variable $\eta$ must be the same as that in the definition of $Sh$ by virtue of \eqref{eq:boundary_condition_liquid}. Indeed, $\partial W/\partial\tZ = F^{\prime} \, \tT^{-\lambda}$. So if the time exponent in $Sh$ and $\eta$ were different, \eqref{eq:boundary_condition_liquid} could not be made self-similar.}

\coltext{Introducing the proposed ansatz into \eqref{eq:simplified_eqn_adv_W}, we get}
\begin{equation}
    \coltext{-\lambda A \tT^{-\lambda-1} - \lambda \eta \tT^{-1}F^{\prime} - \Delta w_i \frac{\rho}{\rho_w} A \tT^{-2\lambda} F^{\prime} = \tT^{-2\lambda}F^{\prime \prime} \, .} 
\end{equation}
\coltext{For $\tT \gg 1$, the first term is always negligible compared to the second one, since $\lambda > 0$. Then, balancing the second term with the last two terms, we obtain that the equation becomes self-similar if $\lambda = 1/2$, as expected from the experiments and the simulations, resulting in}
\begin{equation}
    \coltext{-\left(\frac{1}{2}\eta + A \Delta w_i \frac{\rho}{\rho_w}\right) F^{\prime} = F^{\prime \prime} \, .}
    \label{eq:self_similar_eqn_for_W}
\end{equation}
\coltext{This differential equation can be solved with the boundary conditions $F(0) = 0$ (from \eqref{eq:simplified_eqn_W_at_z0}) and $F^{\prime}(0) = A$ (from \eqref{eq:boundary_condition_liquid}), to yield}
\begin{eqnarray}
    \coltext{F(\eta)} & \coltext{=} & \coltext{-\sqrt{\pi}A\left(1 + w_\mathrm{eq}\frac{\rho}{\rho_w}\right)\exp\left(A^2 \Delta w_i^2 \left(\frac{\rho}{\rho_w}\right)^2\right) \times} \\
    & & \coltext{\left(\mathrm{erf}\left(A \Delta w_i \frac{\rho}{\rho_w}\right)-\mathrm{erf}\left(A \Delta w_i\frac{\rho}{\rho_w} + \frac{\eta}{2}\right)\right) \, .}
    \label{eq:solution_W_eta}
\end{eqnarray}
\coltext{Finally, an additional condition is needed to determine the value of $A$. This condition stems from the behavior of $W$ far away from the evaporation boundary $\tZ = 0$, where $W \rightarrow 1$ (from \eqref{eq:W_short_t} while neglecting $Sh \ll 1$ against unity). Introducing this condition into \eqref{eq:solution_W_eta}, we finally get:}
\begin{equation}
    \coltext{\sqrt{\pi}A \left(1 + w_\mathrm{eq}\frac{\rho}{\rho_w}\right) \exp\left(A^2 \Delta w_i^2 \left(\frac{\rho}{\rho_w}\right)^2\right) \mathrm{erfc}\left(A \Delta w_i \frac{\rho}{\rho_w}\right) - 1 = 0 \, .}
    \label{eq:eq_for_A}
\end{equation}

\begin{figure}
\centering
\includegraphics[scale=1.0]{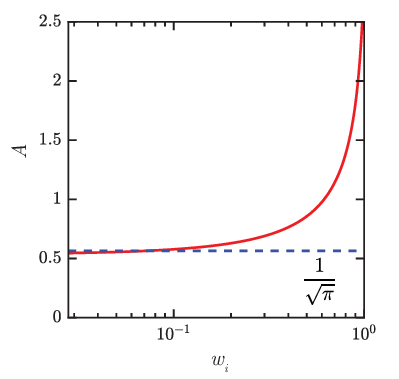}
\caption{\coltext{Coefficient $A$ in $Sh = A \tn^{-1/2}$, obtained by solving (\ref{eq:eq_for_A}) numerically, as a function of the initial water concentration. At low initial water concentration, and under the assumption that $w_\mathrm{eq} \approx 0$, we recover the pure diffusion regime, $A = 1/\sqrt{\pi}$ (blue dashed line). For large water concentrations, $A$ grows unbounded, indicating that the regime $Sh \sim \tn^{-1/2}$ is never reached, as $Sh = 1$ in this limit.}}
\label{fig:A_vs_wi}
\end{figure}

\noindent \coltext{The value of A can be evaluated numerically. Plotting $A$ against $w_i$, we see that $A$ grows monotonically with the initial water concentration, recovering the diffusion-driven asymptotic solution $A = 1/\sqrt{\pi}$ for $w_i \ll 1$ (if the additional assumption that $w_\mathrm{eq} = 0$ is made; figure \ref{fig:A_vs_wi}). This means that higher the initial concentration of water $w_i$, greater is the advective enhancement of mass transfer compared to pure diffusion.}

\subsubsection{\coltext{Uniform approximation}}
\begin{figure}
\centering
\includegraphics[scale=1.0]{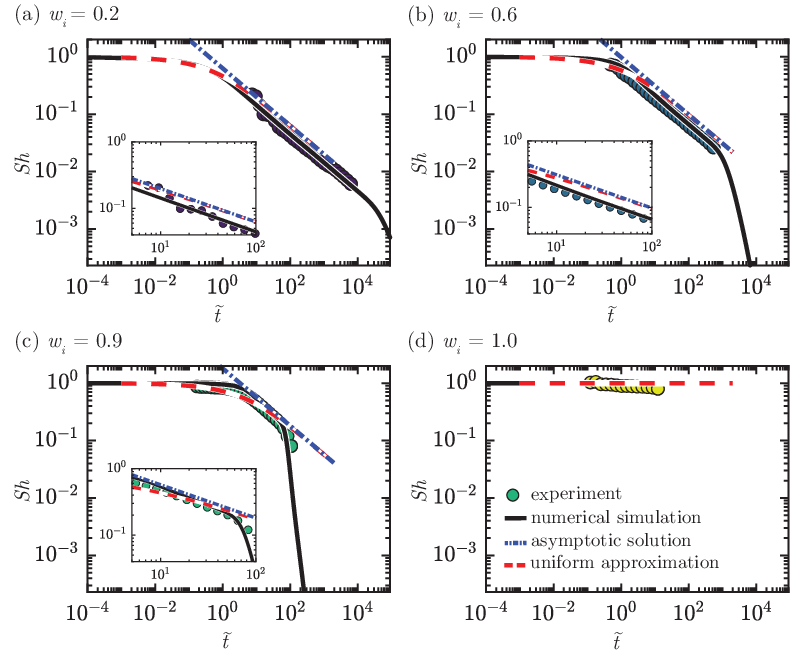}
\caption{Variation of the normalised evaporation rates (Sherwood number, $Sh$) with normalised time ($\tn$) obtained from the experiments (discrete datapoints), the numerical simulations (continuous lines), the asymptotic solution for long times (dash-dotted lines), and the approximate expression (\eqref{eq:Sh_uniform_approx}, dashed lines), for different initial weight fractions of water $w_{i}$: (a) $w_{i} = 0.2$, (b) $w_{i} = 0.6$, (c) $w_{i} = 0.9$, and (d) $w_{i} = 1.0$. \coltext{Insets in panels (a), (b), and (c) highlight the comparisons between the experiments, the numerical simulations, and the simplified analytical modelling.}}
\label{fig:comparison_asymptotic}
\end{figure}

\coltext{For practical applications, it is desirable to have an approximate expression that converges to the asymptotic solutions in the limits $t \ll 1$ ($Sh \approx 1$) and $t \gg 1$ ($Sh \approx~A\tT^{-1/2}$). To this end, we notice that the exact solution for the problem without advection,}
\begin{equation}
    Sh = \exp\left(\tT\right)\mathrm{erfc}\left(\sqrt{\tT}\right),
    \label{eq:ShBimodel1_again}
\end{equation}
\coltext{captures the behavior of the numerical solution with advection, except that the prefactor of the equation $Sh \sim \tT^{-1/2}$ in \eqref{eq:ShBimodel1_again} at $\tn \gg 1$ is $1/\sqrt{\pi}$, instead of $A$ (as in \eqref{eqn:Sh_A_definition}). Thus, we propose the following expression to approximate the full numerical solution uniformly at all times:}
\begin{equation}
    Sh = \exp\left(\frac{\tT}{\pi A^2}\right)\mathrm{erfc}\left(\sqrt{\frac{\tT}{\pi A^2}}\right).
    \label{eq:Sh_uniform_approx}
\end{equation}
\coltext{We observe that this approximation successfully reproduces the numerical simulations much better than the solution without advection, as shown in figure \ref{fig:comparison_asymptotic}.} 

\coltext{To conclude this section, we point out that another uniform approximation to compute the Sherwood number $Sh$, similar to \eqref{eq:Sh_uniform_approx}, can be obtained by solving the simplified problem \eqref{eq:simplified_eqn_adv} - \eqref{eq:simplified_eqn_adv_bc} analytically considering advection as quasi-constant. For completeness, this solution is described in appendix \ref{section:appQuasiAdvectionModel}. The fact that treating advection as quasi-constant yields an expression that works fairly well is interesting, as it points out that advection is mostly important when it is nearly constant, that is, at short times (see figure \ref{fig:PecletNumber}). This makes sense, since it is at this stage when the Peclet number reaches the largest value. This idea could be useful to pursue further analytical approaches for similar problems.}

\subsection{Transient diffusion model with finite length effects} \label{sec:model2}

\coltext{The models described in \S~4.1 and \S~4.2 can faithfully explain both $Sh=1$ and $Sh \sim \tn^{-1/2}$ behaviours seen in the experiments and the simulations. However, we are yet to explain the sharp deviation from $Sh \sim \tn^{-1/2}$ seen for very late times in the simulations at $w_i = 0.6$ and $0.9$ (figures \ref{fig:comparison_asymptotic}b and c). To this end, we turn our attention to the semi-infinite assumption.}

\begin{figure}
\centering
\includegraphics[width=\textwidth]{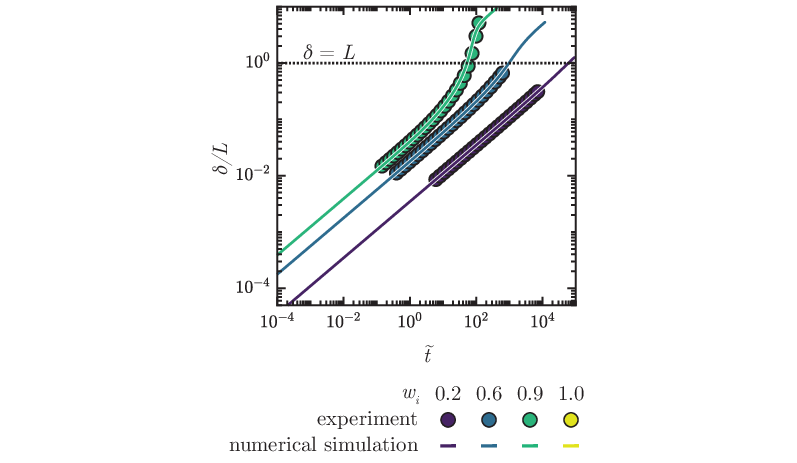}
\caption{ Variation of the penetration depth $\delta$ (normalised with the length $L$ of the liquid column) with the normalised time $\tn$. Discrete datapoints are based on the experiments and continuous lines are based on the numerical simulations. The dotted lines at $\delta/L=1$ indicate when the penetration depth is equal to the size of the liquid column. The penetration depth is defined as $\delta (t) = \sqrt{D t}$ .}
\label{fig:deltaL}
\end{figure}

\coltext{For the semi-infinite assumption to hold, the penetration depth of the diffusion front $\delta (t) = \sqrt{D t}$ should be much smaller than the length of the liquid column $L(t)$. We plot the variation of $\delta/L \left(= \sqrt{D t}/L\right)$ with $\tn$ as obtained from the experiments and the numerical simulations in figure \ref{fig:deltaL} for different $w_i$. It can be observed that for $w_i=0.9$, $\delta/L=1$ at $\tn \approx 60$, which approximately agrees with the time when the slope of $Sh\left( \tn\right)$ starts to deviate from $Sh \sim \tn ^{-1/2}$ in figure \ref{fig:comparison_asymptotic}c. The same holds true for $w_i=0.6$ at $\tn \approx 10^3$ (figure~\ref{fig:comparison_asymptotic}b). Thus, we conclude that although the semi-infinite assumption holds at early times, finite length effects should be included at later times for $w_i=0.6$ and $0.9$ in order to accurately capture the physics of the problem.}

\begin{figure}
\centering
\includegraphics[width=\textwidth]{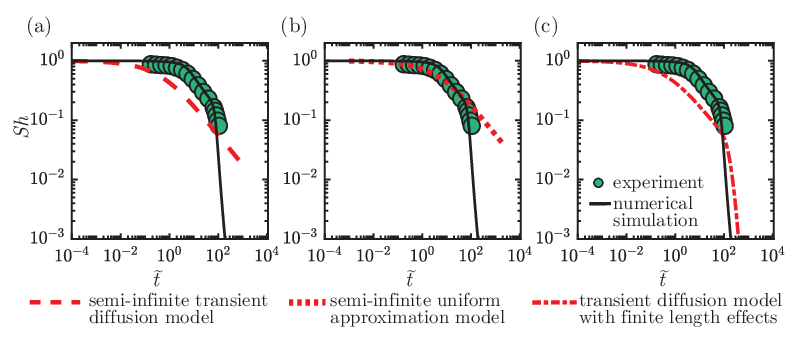}
\caption{ Plots comparing the normalised evaporation rates (Sherwood number, $Sh$) versus normalised time ($\tn$) obtained from the experiments (discrete datapoints) and the numerical simulations (continuous lines) with the theoretical values obtained using the \coltext{(a) semi-infinite transient diffusion model (dashed line), (b) semi-infinite uniform approximation model (dotted line), and (c) transient diffusion model with finite length effects (dash-dotted line) for an initial weight fraction of water $w_i$ = 0.9.}}
\label{fig:ShBiPlots_wo90}
\end{figure}

\coltext{We show in this subsection that this is an effect of the finite length of the capillary, which becomes relevant at long times. To include the effects of the finite length, we use the original boundary conditions of \eqref{eqn:bc_main}.} 

\coltext{We will first check how the three terms in the governing equation \eqref{eqn:neq_c0}, reproduced here for convenience, compare with each other for the late times when the finite length effects start playing a role:} 

\begin{equation}
    \frac{\partial\tilde{w}}{\partial\tilde{t}} - \frac{\rho \Delta w_i}{\rho_w}\,Sh\,\frac{\partial\tilde{w}}{\partial\tilde{z}} = \frac{\partial^2\tilde{w}}{\partial\tilde{z}^2} \, .
    \label{eqn:neq_c0_again}
\end{equation}

\coltext{When the boundary layer becomes of the order of the length of the domain, $\delta \sim L$, the first and second spatial derivatives in \eqref{eqn:neq_c0_again} scale as $\Delta\wn/\Ln$ and $\Delta\wn/\Ln^2$, respectively. For $w_i < 1$, the volume occupied by water at long times is small compared to that occupied by glycerol, so $\Ln$ tends asymptotically to a constant as the total liquid volume changes very slowly.
This means that the order of magnitude of the first and second spatial derivatives differ in a constant factor $\Ln$. In this situation, since the prefactor of the advective term goes to zero as $Sh \rightarrow 0$, the advective term is going to be a factor $Sh$ smaller than the diffusive one and can be neglected.
Thus, we formulate a quasi-constant length transient diffusion model valid for $Sh \ll 1$.}

\coltext{The governing equations and the initial and boundary conditions (\eqref{eqn:neq_c0} - \eqref{eqn:Sh_w_relation}) can now be written as}  
\begin{equation}
    \frac{\p \wn}{\p \tn} = \frac{\p^2 \wn}{\p \zn^2} \, ,
    \label{eqn:neq_c2}
\end{equation}

\begin{equation}
 \begin{array}{cl}
\displaystyle\wn = 0 
    &\quad \mbox{at\ }\quad \tn=0\, ,\\[8pt]
\displaystyle\frac{\p \wn}{\p \zn} = 0 
    &\quad \mbox{at\ }\quad \zn = \Ln \, \, \\[8pt]
\displaystyle \frac{\p \wn}{\p \zn} = \wn-1 
    &\quad \mbox{at\ }\quad \zn=0 \, ,
 \end{array}
  \label{nbc_c2}
\end{equation}

\begin{equation}
    \coltext{Sh = 1 - \wn(\zn=0, \tn) \, .}
    \label{eqn:Sh_w_relation3}
\end{equation}

\noindent The solution of this system of equations is given by \citep{Crank1975}:
\begin{equation}
    \wn(\zn,\tn, \Ln) = 1 - \sum_{n=1}^{\infty} \left( \frac{2 \Ln \, \cos \left( \lambda_n \left( 1 - \zn / \Ln \right)\right) \exp \left(-\lambda_n^2 \, \tn / \Ln^2\right)}{\left( \lambda_n^2 + \Ln^2 + \Ln \right) \cos(\lambda_n)} \right) \, ,
    \label{eqn:wnFinal_c2}
\end{equation}

\begin{equation}
   \coltext{ Sh = 1 - \tW(\tZ=0, \tT) = \sum_{n=1}^\infty \frac {2\Ln \, \exp \left( -\lambda_n^2 \, \tn / \Ln^2 \right)}{\left( \lambda_n^2 + \Ln^2 + \Ln \right)} \, ,}
    \label{eqn:ShBimodel2}
\end{equation}

\noindent \coltext{where $\lambda_n$ is the $n^{\text{th}}$ root of the equation $\lambda \, \tan (\lambda) = \Ln$. Combining \eqref{eqn:ShDefine} and \eqref{eqn:ShBimodel2}, the velocity of the top interface can now be written as}

\begin{equation}
    \frac{\mathrm{d}L}{\mathrm{d}t} = v_y = \frac{h^{\ast} \Delta w_i}{\rho_w}   \sum_{n=1}^\infty \frac{2\Ln \, \exp \left( -\lambda_n^2 \, \tn / \Ln^2 \right)}{\left( \lambda_n^2 + \Ln^2 + \Ln \right)} .
    \label{eqn:vtmodel2}
\end{equation}

Equation \eqref{eqn:vtmodel2} is integrated in time using the built-in Matlab function \texttt{ode45} (which implements an adaptive-time-step Runge-Kutta algorithm of fourth order) to obtain $L(t)$, and subsequently the variation of $Sh$ with $\tn$ (figure \ref{fig:ShBiPlots_wo90}c). For $\tn \gg 1$, $v_y$ can be approximated by the first term in the series:
\begin{equation}
    v_y = \frac{h^{\ast} \Delta w_i}{\rho_w} \frac{2\Ln \, \exp \left( -\beta_1^2 \, \tn / \Ln^2 \right)}{\left( \beta_1^2 + \Ln^2 + \Ln \right)} .
    \label{eqn:vt_1term}
\end{equation}
When the changes in the length $L$ of the liquid column are slow, the velocity decreases exponentially with time (as per \eqref{eqn:vt_1term}), thus correctly predicting the late time $Sh (\tn)$ behaviour of $w_i=0.9$ in the simulations (figure \ref{fig:ShBiPlots_wo90}c). Note that observing this regime experimentally is complicated from the practical point of view, due to the very small velocities associated with it. Although we can observe it easily in the numerical simulations, we are only able to see the beginning of this regime in the experiments for the most favourable case of $w_i = 0.9$, for which this regime appears at velocities of the order of less than a micron per second (see figure \ref{fig:ytvt_Plots}). \coltext{In conclusion, this model successfully captures the deviation from $Sh\sim \tn^{1/2}$ due to the finite length effects.}

\section{Conclusions and outlook} \label{sec:conclusion}

In this work, we studied the evaporation of aqueous glycerol solutions in cylindrical capillaries having a circular cross section. We characterised the drying behaviour in terms of a normalised mass transfer rate (Sherwood number $Sh$) and a normalised time $\tn$. Our experiments quantitatively demonstrate how the addition of glycerol reduces the evaporation rate of water. The corresponding direct numerical simulations indicate that modelling the system as a one-dimensional advection-diffusion mass transfer problem with composition-dependent properties can quantitatively reproduce the evaporation behaviour observed in the experiments. The evaporation of water shows three main regimes: (i) $Sh = 1$, (ii) $Sh \sim 1/\sqrt{\tn}$, and (iii) $Sh \sim \exp\left(-\tn\right)$. We describe the physical origins of these regimes using a one-dimensional simplistic analytical model with constant material properties and a linearised composition-dependent activity of water.

Modelling the system as a problem of pure diffusion in a semi-infinite medium reproduces $Sh = 1$ and $Sh \sim 1/\sqrt{\tn}$ as the early time and late time behaviours, respectively. $Sh=1$ in the early time limit corresponds to a rapid replenishment of water at the evaporating interface, leading to a constant evaporation rate and constant interfacial concentration of water. In the late time regime, replenishment of the interfacial concentration of water is limited by diffusion, leading to the classical diffusion-like $Sh \sim 1/\sqrt{\tn}$ behaviour. 

\coltext{However, we also show that even if the pure diffusion model captures the scaling relations of $Sh(\tn)$ correctly, advective replenishment of water needs to be considered for more precise prediction of the evaporation rates. A Peclet number defined as $\Delta w_i \, Sh \, \rho / \rho_w = \rho \, v_y / h^{\ast} = (w \vert_{z=0} - w_\mathrm{eq}) \rho/\rho_w$ (the coefficient of the advective term in \eqref{eqn:neq_c0}) is the relevant parameter dictating the importance of advection. Thus, advection is small when the interfacial concentration of water $w \vert_{z=0}$ is close to the equilibrium concentration $w_{\mathrm{eq}}$, and high otherwise.} 

\coltext{Finally, we show that as the diffusive penetration depth $\delta = \sqrt{Dt}$ increases and the length $L(t)$ of the system decreases, they can become comparable in magnitude. In such a scenario, the semi-infinte approximations hold only as long as the late time behaviour is modified to $Sh \sim \exp(-\tn)$. This change in the evaporation regime essentially reflects the effect of the finite size of the liquid column.}

Even though a model with constant material properties was used to describe the evaporation of a binary mixture, the spatio-temporal changes in properties such as the density and the diffusion coefficient affect the precise quantitative prediction of evaporation rates. The direct numerical simulations are devoid of these deficiencies. However, the simplified analytical models provide valuable insight into the essential physics of the system and can provide predictions for more complex liquid mixtures.

We also note that for predicting the evaporation rate, we have used the expression corresponding to that of a thin droplet on a substrate ($\theta_\text{drop}=0^\circ$), giving excellent predictions that match the experimental observations. However, one can include corrections to the mass transfer coefficient $h^{\ast}$ to account for the difference between the air-liquid interface of a droplet and the air-liquid interface at the mouth of a capillary tube \citep{Junhui2019}. Moreover, during evaporation, the shape of the lower interface changes ($\theta<90^\circ$) until the meniscus depins. This change in the shape of the interface might also require a small correction to $h^{\ast}$, and might explain the very small decrease in $Sh$ seen in the experiments of pure water (figure \ref{fig:ytvt_Plots}b). However, further studies along the lines of \citet{Ambrosio_2021} are required to confirm this hypothesis.

Finally, all measurements were limited to times during which the lower meniscus was pinned at the mouth of the capillary. When the lower meniscus depins and propagates into the capillary, the rate-limiting step of evaporation would change from 3-D vapour diffusion to 1-D vapour diffusion. In the case of a single component liquid evaporating from a square capillary, there have been efforts to predict the time when the lower meniscus depins \citep{Chauvet2010}. Further studies are required to predict the same for multi-component liquids and capillaries of various geometries and configurations (e.g. inclination with respect to gravity). Nonetheless, in this study, we have shown that these kind of phenomena may not be essential to predict, with a reasonable degree of accuracy, the mass transfer rate in a relatively complex system like the one we consider here.

The aforementioned results can also be directly applied to predict the evaporation of multi-component liquids from porous structures, which can be modelled as bundles of thin capillaries. Evaporation from capillaries can also help us understand the evaporative behaviour of biological fluids (such as blood, saliva, or liquids in respiratory droplets \citep{OmerPNAS2022, Seyfert2022}) or novel liquid mixtures (for applications such as evaporative cooling and spray drying). Lastly, studying evaporation from capillaries can also be useful in the case of inkjet printing, where the evaporation from the tip of the printing nozzle can lead to changes in the composition of the ink \citep{rump2022selective}. Our model can provide insight into the changes in the composition at the nozzle tip for a given time scale and assist in carefully choosing the properties of the ink.

\noindent{\textbf{Acknowledgments}}

The authors thank Andrea Prosperetti for insightful discussions, and Martin Bos, Gert-Wim Bruggert, Dennis van Gils, and Thomas Zijlstra for their assistance with the fabrication of the controlled humidity chamber. 

\noindent{\textbf{Funding}}
 
We acknowledge the funding by an Industrial Partnership Programme of the Netherlands Organisation for Scientific Research (NWO), co-financed by Canon Production Printing B. V., University of Twente, and Eindhoven University of Technology. D.L. acknowledges funding by the ERC Advanced Grant No. 740479-DDD. J.R.R. acknowledges funding from the Spanish MCIN/AEI/10.13039/501100011033 through grant no. PID2020-114945RB-C21. X.H.Z. acknowledges the support
by the Natural Sciences and Engineering Research Council of Canada (NSERC), Alberta Innovates, and the support from the Canada
Research Chairs program.\\

\noindent{\textbf{Declaration of interests}}

The authors report no conflict of interest. \\

\noindent{\textbf{Supplementary information}} \label{SM}

Supplementary information is available at (URL to be inserted by publisher). \\

\noindent{\textbf{Author ORCID}}

L. Thayyil Raju \href{https://orcid.org/0000-0002-2054-3884}{https://orcid.org/0000-0002-2054-3884}; 

C. Diddens \href{https://orcid.org/0000-0003-2395-9911}{https://orcid.org/0000-0003-2395-9911};

J. Rodríguez-Rodríguez \href{https://orcid.org/0000-0001-8181-138X}{https://orcid.org/0000-0001-8181-138X};

X. Zhang \href{https://orcid.org/0000-0001-6093-5324}{https://orcid.org/0000-0001-6093-5324}

D. Lohse \href{https://orcid.org/0000-0003-4138-2255}{https://orcid.org/0000-0003-4138-2255};

U. Sen \href{https://orcid.org/0000-0001-6355-7605}{https://orcid.org/0000-0001-6355-7605} .\\

\appendix

\section{Evaporation from the upper interface}\label{section:appA}

The evaporative flux from the top interface is given by \citep{Stefan1873}:
\begin{equation}
    \frac{\mathrm{d}M^{\prime \prime}}{\mathrm{d}t} = \frac{D_{v,a}}{L} \frac{p}{RT} \ln \left(\frac{p - p_{w,L}}{p - p_{w,s}} \right) ,
    \label{eqn:topEvap1}
\end{equation}
where $p$ is the atmospheric pressure, $p_{w,s}$ the partial pressure of water vapour at the surface of the upper meniscus, $p_{w,L}$ the partial pressure of water vapour at the upper end of the capillary tube, and $L$ the length of the capillary tube above the upper meniscus. Since $p_{w,s}$ is much smaller than $p$, the above expression can be simplified as 
\begin{equation}
    \frac{\mathrm{d}M^{\prime \prime}}{\mathrm{d}t} = \frac{D_{v,a}}{L} \left(\frac{p_{w,s}}{RT} - \frac{p_{w,L}}{RT}\right) = \frac{D_{v,a}}{L} \left(c_{w,s}-c_{w,L}\right) .
    \label{eqn:topEvap2_P}
\end{equation}
Thus the ratio of evaporative flux from the top interface to the bottom interface can be estimated as 
\begin{equation}
    \frac{\frac{\mathrm{d}M_{\text{top}}^{\prime \prime}}{\mathrm{d}t}} {\frac{\mathrm{d}M_{\text{bottom}}^{\prime \prime}}{\mathrm{d}t}} = \frac{\frac{D_{v,a}}{L} \left(c_{w,s}-c_{w,L}\right)_{\text{top}}}{\frac{4D_{v,a}}{\pi R} \left(c_{w,s}-c_{w,L} \right)_{\text{bottom}}} \, .
    \label{eqn:compareTopBottom}
\end{equation}
Thus, even if the top of the capillary is subjected to the same humidity as the bottom, the evaporation from the top is lower by a factor of $\pi R/4L=0.005$. Hence, the evaporation from the upper interface can be neglected. 

\section{Spatio-temporal changes in the concentration distribution within the capillary}\label{section:appAxialConc}
\begin{figure}
\centering
\includegraphics[width=\textwidth]{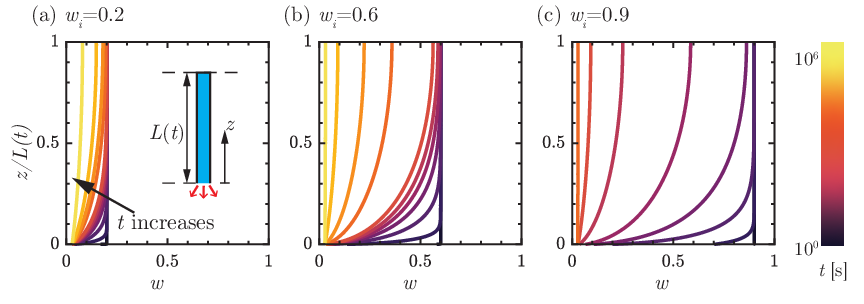}
\caption{Axial distribution of the concentration field for three different initial concentrations: (a) $w_{i} = 0.2$, (b) $w_{i} = 0.6$, and (c) $w_{i} = 0.9$. The colors represent different time stamps: $1 \times 10^{0}$~s, $1 \times 10^{3}$ s, $1 \times 10^{4}$ s, $3 \times 10^{4}$ s, $5 \times 10^{4}$ s, $7 \times 10^{4}$ s, $1 \times 10^{5}$ s, $3 \times 10^{5}$ s, $5 \times 10^{5}$ s, $1 \times 10^{6}$ s, and $3 \times 10^{6}$ s.}
\label{fig:axialDistribution}
\end{figure}

We show in figure \ref{fig:axialDistribution} snapshots of the concentration field inside the capillary for different initial water concentrations computed using the full numerical simulations described in \S~\ref{sec:numerics}.
\FloatBarrier

\section{Linearisation of the water vapour concentration difference}\label{section:app_Linearisation}

\begin{figure}
\centering
\includegraphics[width=\textwidth]{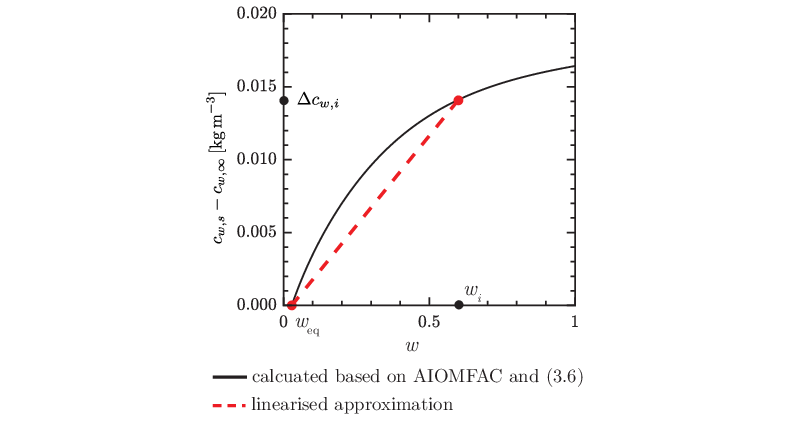}
\caption{Difference in concentration of water vapour at the air-liquid interface and far away from interface, at a temperature of $21^\circ$C and relative humidity $H_r = 10 \%$. The solid black line is calculated using AIOMFAC \citep{Zuend2011} and \eqref{eqn:cws}. The dotted red line shows the linearised approximation used in \eqref{eqn:deltaCs_linear} in the analytical model for $w_i=0.6$. $w_\mathrm{eq}=0.028$ for our case}
\label{fig:deltaC}
\end{figure}

The difference in concentration $c_{w,s} - c_{w,\infty}$ of water vapour is a nonlinear function of the weight fraction of water $w$ (figure \ref{fig:deltaC}; based on \eqref{eqn:cws}). For our analytical model, we linearise $c_{w,s} - c_{w,\infty}$ between the initial concentration $w_i$ and the equilibrium concentration $w_\mathrm{eq}$. In figure \ref{fig:deltaC}, the corresponding coordinates are $(w_i,\Delta c_{w,i})$ and $(w_\mathrm{eq},0)$, where

\begin{equation}
    \Delta c_{w,i} = c_{w,s} - c_{w,\infty},
\end{equation}
\noindent is the initial difference in the concentration of water vapour. From \eqref{eqn:cws}, and the definition of $c_{w,\infty}=H_r c_{w,s}^o$, we get 

\begin{equation}
    \Delta c_{w,i} = a \, c_{w,s}^o - H_r c_{w,s}^o = x_{i} \, \psi_{i} \, c_{w,s}^o - H_r c_{w,s}^o .
\end{equation}

\noindent The equation for the line passing through these two points is given by 

\begin{equation}
    \frac{\left(c_{w,s} - c_{w,\infty}\right) - 0}{\left(x_{i} \, \psi_{i} \, c_{w,s}^o - H_r c_{w,s}^o\right) - 0} = \frac{w - w_\mathrm{eq}}{w_i - w_\mathrm{eq}} ,
\end{equation}
which can be rewritten as 
\begin{equation}
    c_{w,s} - c_{w,\infty} = c_{w,s}^0 \, \frac{x_i \, \psi_i - H_r}{\Delta w_i} \,  (w - w_\mathrm{eq}) \, .\label{eqn:deltaCs_linear_appendix}
\end{equation}

Since \eqref{eqn:deltaCs_linear_appendix} applies at the boundary $z=0$, we replace $w$ in the right hand side of \eqref{eqn:deltaCs_linear_appendix} with $w \vert_{z=0}$ to get \eqref{eqn:deltaCs_linear}.

\section{Properties of water-glycerol mixtures}\label{section:app_property}

\begin{figure}
\centering
\includegraphics[width=\textwidth]{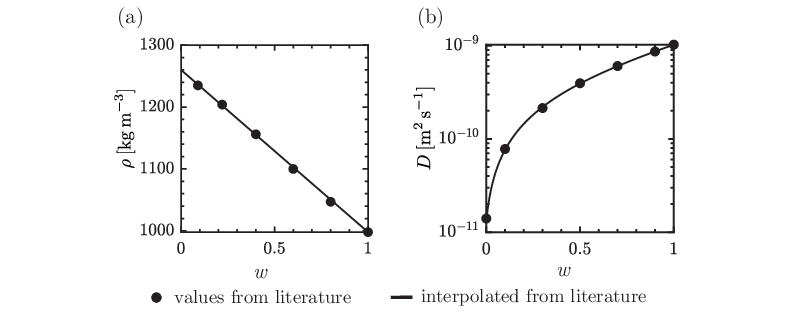}
\caption{Material properties of the glycerol-water mixtures used in our direct numerical simulations. Data taken from \citet{DErrico2004} and \citet{Takamura2012}.}
\label{fig:materialProp}
\end{figure}

We include in this appendix plots showing the dependency of the density and diffusivity of water-glycerol mixtures as a function of the mass fraction of water. These curves are the ones used in the numerical simulations described in \S~\ref{sec:numerics}. For the analytical models described in \S~\ref{sec:modellingStart}, we used constant values corresponding to the initial water concentration, which can also be read from figure \ref{fig:materialProp}.

\section{\coltext{Simplified model with quasi-constant advection}}\label{section:appQuasiAdvectionModel}

In this section, we will elaborate on the quasi-constant advection model referred to at the end of \S~\ref{sec:model3}. \coltext{In addition to the approximations made at the beginning of \S~\ref{sec:modellingStart}, we assume that the changes in velocity (and thus in $Sh$) are slow compared to the response of the concentration field, such that velocity (and thus $Sh$) can be considered quasi-constant for the purpose of determining the concentration field.} Thus, we develop a transient diffusion model with quasi-constant advection. 

We use the full form of \eqref{eqn:neq_c0} with the initial and boundary conditions given in \eqref{eqn:bc_main}, except that we modify the boundary condition at $\zn = \Ln$:

\begin{equation}
    \frac{\p \wn}{\p \tn} - \frac{\rho \Delta w_i Sh}{\rho_w} \, \frac{\p  \wn}{\p \zn} = \frac{\p^2 \wn}{\p \zn^2} ,
    \label{eqn:neq_c4}
\end{equation}

\begin{equation}
 \begin{array}{cl}
\displaystyle\wn = 0 
    &\quad \mbox{at\ }\quad \tn=0,\\[8pt]
\displaystyle\frac{\p \wn}{\p \zn} = 0 
    &\quad \mbox{at\ }\quad \zn \rightarrow \infty, \\[8pt]
\displaystyle \frac{\p \wn}{\p \zn} - \left(1 - \frac{\rho \Delta w_i Sh}{\rho_w}\right) \, \wn + \left( 1 - \frac{\rho w_i Sh}{\rho_w}\right) = 0
    &\quad \mbox{at\ }\quad \zn=0 \, .
 \end{array}
  \label{eqn:nbc_c4}
\end{equation}

We start with the Laplace transform of \eqref{eqn:neq_c4} and \eqref{eqn:nbc_c4}, to obtain
\begin{equation}
    s \wh - \wn(\tn=0) - \alpha \frac{\p \wh}{\p \zn} = \frac{\p^2 \wh}{\p \zn^2} ,
\end{equation}

\begin{equation}
 \begin{array}{cl}
\displaystyle\frac{\p \wh}{\p \zn} = 0 
    &\quad \mbox{at\ }\quad \zn \rightarrow \infty, \\[8pt]
\displaystyle \frac{\p \wh}{\p \zn} - (1-\alpha) \wh + \frac{\beta}{s} = 0 
    &\quad \mbox{at\ }\quad \zn=0 ,
 \end{array}
  \label{lbc_c3}
\end{equation}
where $\wh$ is the Laplace transform of $\wn$ and $s$ is the variable in the Laplace domain. For the sake of brevity, we define

\begin{equation}
    \alpha = \frac{\rho_w \Delta w Sh}{\rho},
    \label{eqn:alphaDefine}
\end{equation}

\noindent and 

\begin{equation}
    \beta = 1 -  \frac{\rho_w w_i Sh}{\rho}\, .
    \label{eqn:betaDefine}
\end{equation}

\noindent Solving this system of equations, we get  
\begin{equation}
    \wh = \frac{-\beta}{s} \frac{1}{\frac{\alpha}{2}-1-\sqrt{s+\frac{\alpha^2}{4}}} \exp \left( -\frac{\alpha \zn}{2} - \zn \sqrt{\frac{\alpha^2}{4}+s}\right) .
\end{equation}
Taking the inverse Laplace transform, we arrive at
\begin{equation}
        L^{-1}\left(\wh\right) = \wn = \beta \exp \left(\frac{-\alpha \zn}{2}\right)
        \, L^{-1} \left(
          \frac{1}{s} \, \frac{1}{\frac{\alpha}{2}-1-\sqrt{s+\frac{\alpha^2}{4}}} \exp \left( -\zn \sqrt{\frac{\alpha^2}{4}+s}\right) \right) .
\end{equation}
Using the identity 
\begin{equation}
    L^{-1}[F(s-\zeta)] = \exp (\zeta) L^{-1} F(s),
\end{equation}
with $\zeta = -\alpha^2/4$, we obtain
\begin{equation}
    \wn = \beta \, \exp \left(\frac{-\alpha \zn}{2}\right) \exp \left(\frac{-\alpha^2\tn}{4}\right) \, L^{-1} \left(\frac{1}{s-\frac{\alpha^2}{4}} \, \frac{1}{1-\frac{\alpha}{2}+\sqrt{s}} \, \exp \left( -\zn \sqrt{s}\right) \right) .
\end{equation}
Further, we can write \citep{1982SolutionList}
\begin{equation}
    L^{-1} \left(\exp \left( \frac{-x\sqrt{s}}{(s-\mu^2)(\xi+s)}\right)\right) = \frac{C}{2(\mu+\xi)} - \frac{D}{2(\mu-\xi)} + E ,
\end{equation}
where
\begin{equation}
    C = \exp (\mu^2t - \mu x) \, \mathrm{erfc}\left(\frac{x}{2\sqrt{t}} - \mu\sqrt{t}\right) ,
\end{equation}
\begin{equation}
    D = \exp (\mu^2t + \mu x) \, \mathrm{erfc}\left(\frac{x}{2\sqrt{t}} + \mu\sqrt{t}\right) ,
\end{equation}
and
\begin{equation}
    E = \frac{\xi}{\mu^2-\xi^2} \, \exp(\xi^2t+\xi x) \, \mathrm{erfc}\left( \frac{x}{2\sqrt{t}} + \xi \sqrt{t}\right) ,
\end{equation}
where having $\mu=\alpha/2$ and $\xi=1 - \alpha/2$, we get
\begin{equation}
    \wn = F + G + H ,
    \label{eqn:wnFinal_c4}
\end{equation}
where
\begin{equation}
    F = \frac{\beta}{2} \exp (-\alpha\zn) \mathrm{erfc} \left( \frac{\zn}{2\sqrt{\tn}} - \frac{\alpha \sqrt{\tn}}{2}\right) ,
\end{equation}
\begin{equation}
    G = - \frac{\beta}{2(\alpha-1)} \mathrm{erfc} \left( \frac{\zn}{2\sqrt{\tn}} + \frac{\alpha \sqrt{\tn}}{2}\right),
\end{equation}
and 
\begin{equation}
    H = \frac{\beta}{2} \frac{2-\alpha}{\alpha-1} \, \exp \left( (1-\alpha) \tn \right) \, \exp \left( (1-\alpha) \zn \right) \, \mathrm{erfc} \left( \frac{\zn}{2\sqrt{\tn}} + \left(1-\frac{\alpha}{2} \right) \sqrt{\tn}\right) .
\end{equation}

\noindent Finally, we use \eqref{eqn:Sh_w_relation} along with \eqref{eqn:wnFinal_c4}, to obtain an implicit equation for the Sherwood number,
\begin{equation}
\begin{split}
    Sh &= 1 - \tW(\tZ=0, \tT) \\ 
    &= 1 - \beta \left( 1 + \frac{\alpha}{2(1-\alpha)} \mathrm{erfc}\left(\frac{\alpha\sqrt{\tn}}{2} \right) - \frac{2-\alpha}{2(1-\alpha)} \exp ((1-\alpha)\tn) \, \mathrm{erfc}\left( \left(1-\frac{\alpha}{2}\right)\sqrt{\tn} \right) \right) \, ,
    \label{eqn:ShBimodel_old3}
\end{split}
\end{equation}

with

\begin{equation}
    \alpha = \frac{\rho_w \Delta w Sh}{\rho} \, ,
    \label{eqn:alphaDefine2}
\end{equation}

\noindent and 

\begin{equation}
    \beta = 1 -  \frac{\rho_w w_i Sh}{\rho}\, .
    \label{eqn:betaDefine2}
\end{equation}

\begin{figure}
\centering
\includegraphics[scale=1.0]{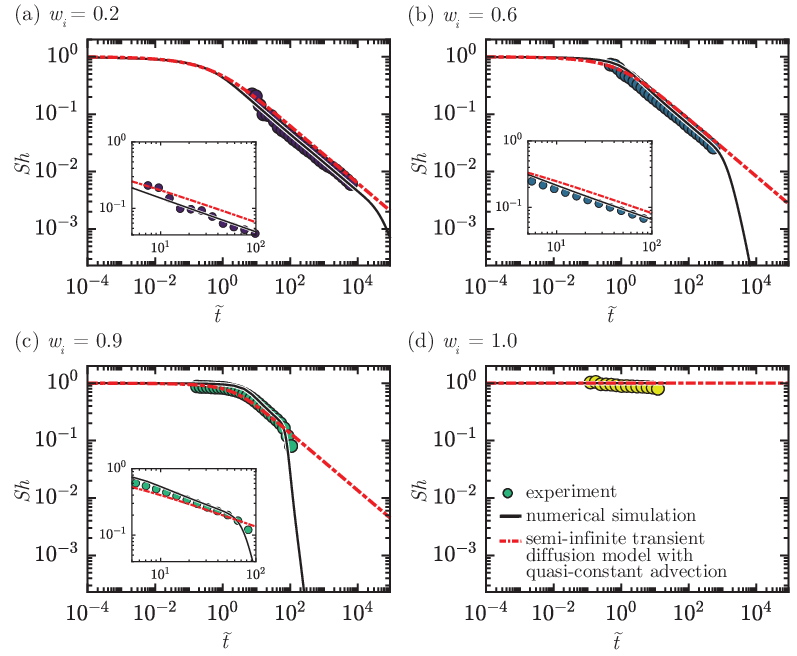}
\caption{Variation of the normalised evaporation rates (Sherwood number, $Sh$) with normalised time ($\tn$) obtained from the experiments (discrete datapoints) and the numerical simulations (continuous lines) with the theoretical values (dashed lines) obtained from the semi-infinite transient diffusion model with quasi-constant advection, for different initial weight fractions of water $w_i$: (a) $w_{i} = 0.2$, (b) $w_{i} = 0.6$, (c) $w_{i} = 0.9$, and (d) $w_{i} = 1.0$. \coltext{Insets in panels (a), (b), and (c) highlight the comparisons between the experiments, the numerical simulations, and the simplified analytical modelling.}}
\label{fig:ShBiPlots_advection}
\end{figure}

\noindent \coltext{Equation \eqref{eqn:ShBimodel_old3} must be interpreted as an implicit algebraic equation to obtain $Sh$. The results of the quasi-constant advection with transient diffusion model \eqref{eqn:ShBimodel_old3} are shown in figure \ref{fig:ShBiPlots_advection}, which indeed demonstrates a reasonably good quantitative agreement with the experiments and the direct numerical simulations. A close examination of \eqref{eqn:ShBimodel_old3} reveals that in the early time limit ($\tn \rightarrow 0$), we get $Sh \rightarrow 1$. Conversely, in the late time limit ($\tn \gg 1$), we recover the diffusion-driven law $Sh \sim A \tn^{-1/2}$.}  

\bibliographystyle{jfm}
\bibliography{CapillaryEvaporation_Ref_v20221026}

\end{document}